# 2060: Civilization, Energy, and Progression of Mankind on the Kardashev Scale

Antong Zhang[1], Jiani Yang[1,*], Yangcheng Luo[1], Siteng Fan[2,**]

[1]Division of Geological and Planetary Sciences, California Institute of Technology, Pasadena, CA 91125, USA

[2]LMD/IPSL, Sorbonne Université, PSL Research University, École Normale Supérieure, École Polytechnique, CNRS, Paris 75005, France

*Correspondence: yjn@caltech.edu (J.Y.); sfan@lmd.ipsl.fr (S.F.)

**ABSTRACT**

Energy has been propelling the development of human civilization for millennia, and technologies acquiring energy beyond human and animal power have been continuously advanced and transformed. In 1964, the Kardashev Scale was proposed to quantify the relationship between energy consumption and the development of civilizations. Human civilization presently stands at Type 0.7276 on this scale. Projecting the future energy consumption, estimating the change of its constituting structure, and evaluating the influence of possible technological revolutions are critical in the context of civilization development. In this study, we use two machine learning models, random forest (RF) and autoregressive integrated moving average (ARIMA), to simulate and predict energy consumption on a global scale. We further project the position of human civilization on the Kardashev Scale in 2060. The result shows that the global energy consumption is expected to reach 928-940 EJ in 2060, with a total growth of over 50% in the coming 40 years, and our civilization is expected to achieve Type 0.7474 on the Kardashev Scale, still far away from a Type 1 civilization. Additionally, we discuss the potential energy segmentation change before 2060 and present the influence of the advent of nuclear fusion in this context.

**INTRODUCTION**

Throughout the history of human civilization, energy has been holding an imperative role in humanity's progress.[1] Especially in the past few centuries, innovations in the harnessing of power have catalyzed humanity's rapid growth.[2] In the 18th century, the advent of the Industrial Revolution led to significant technological developments. The powering of steam engines that consume fossil fuels increased the production in factories by orders of magnitude[3]. The evolution of energy continues to be the key to humanity's development.[2,4] Giving complete rein of human beings over nature, a series of similar industrial and agricultural revolutions betokened the future of mankind's reliance on energy. As the advancement progresses, the invention of electricity further opened mankind's eyes to a promising future with energy.[5] Humanity has hitherto stayed true to the energy demand and consumption principle, growing at a compound annual growth rate of 2.43% from 1965-2020.[6,7] With humanity progressing at a remarkable rate in the past few centuries, however, the pace at which human beings could continuously progress as a civilization in the future remains uncertain.

As mankind establishes its identity in the universe, insatiable human curiosity over the realm of civilization reaches its peak in the 1960s,[8] which conducted mankind to cogitate more deeply into the concept of civilization. Providing that some of the extraterrestrial civilizations are highly likely million years more advanced than mankind, Soviet astrophysicist Nikolai Kardashev proposed a scale, which is later known as the Kardashev Scale, to classify a civilization's technological development based on its energy consumption.[9] The scale initially categorized civilizations into three types. Type 1 is known as the planetary civilization, which features the capability of harnessing and utilizing all forms of energies that can be reached on the host planet, such as fossil fuels, solar, and geothermal power. Similarly, Type 2 and 3, known as the stellar and galactic civilizations, respectively, are capable of extracting and utilizing all energy created by their respective systems.[9] The scale proved lackluster in quantitative presentation of the civilization types. Subsequently, Carl Sagan furthered the Kardashev Scale by extrapolating data, and proposed a continuous function quantifying the Kardashev Scale in index K[10]

$$K = \frac{\log P - 6}{10}$$

where P represents energy consumption rate in Watt. Sagan estimated that, by approximation, a Type 2 civilization should meet an energy consumption rate of $10^{26}$ W and a Type 3 civilization of $10^{36}$ W, both of which represent the cumulative energy output of their respective systems. Extrapolating these two values, he suggested a Type 1 civilization to have an energy consumption rate of $10^{16}$ W.

There is much progress to be desired before humanity is able to acquire the energy capacity to take its first stride on the Kardashev Scale. Presently, mankind is measured on the scale at K = ~0.7276;[7] humanity is but a mere sapling that has just budded. On the bright side, the plethora of promising new technologies being developed and others that are yet to come betoken rapid progress. Given the current energy structure of human civilization, to proceed on the Kardashev Scale, it is imperative to replace the anachronous, unsustainable methods of energy extraction with more efficient and renewable ones. Consisting of an alarming 80% of current energy consumption rate[11], fossil fuels have proved and will prove pernicious to humanity's future.[12] Renewable energy sources are frequently mentioned as the most promising substitute for fossil fuels;[13] however, despite its undeniable advantages, clean energy technology will not be able to take dominance over fossil fuels in at least the few coming decades due to its immatureness as well as some known, unavoidable limitations.[14,15] Another potentially more promising alternative that may behoove mankind is nuclear fusion. It has a set of unique advantages and an ability to generate nearly inexhaustible, pollution-free energy, which has allowed humanity to see the light at the end of the tunnel.[16] The world's largest fusion experiment ITER anticipates the first full-scale plasma in 2025, and the testing for fusion is expected to start a decade later.[17] Through a testing period of ~25 years, the European roadmap targets to generate net energy output and establish facilities to convert said energy to electricity.[18] Therefore, along with other promises and predictions of the energy future, nuclear fusion has the potential to prompt the role of the 6th decade of this century as a milestone of humanity's energy progression and transformation.[19-21]

The development of the Kardashev Scale and the Drake Equation[22] inspired continuous interest in the search for technologically developed extraterrestrial intelligence. However, quantitative predictions of our own civilization, humanity, have rarely been touched. A recent study using linear regression predicted the trend of K value and concluded that human civilization is projected to reach Type 1 in the mid-24th century.[23] Owing to the complexity of the energy structure and different future of energy sources, applying linear regression onto one univariate time series oversimplifies the relationship between total energy consumption and the main contributors. Furthermore, this study assumed a linear trend of index K as the prediction parameter rather than energy consumption, but the historical data of K presents a logarithmic trend, and limitations are significant due to the lack of consideration for several key influencing factors. To resolve this issue, methods that are capable of dealing with high complexity are required to arrive at a precise and reliable conclusion.

As a result of the development of computing power, the advent of artificial intelligence (AI) and machine learning (ML) approaches have been improving our capability of energy forecasting in prodigious ways. ML models are typically suitable for finding patterns or hidden information in historical data and making predictions, as long as no substantial changes take place in the composition structure. In the past, deep neural networks have been extensively used to predict energy consumption[24] due to its performance for making forecasts in terms of coefficient of determination ($R^2$) and root mean square error (RMSE). However, there have been existing difficulties to evaluate the influence of individual input variable.[23,25] Besides neural networks, the random forest[26,27] (RF) model is an ensemble learning algorithm for classification, which yields reliable results using predictions derived from decision trees.[28] RF is also capable of dealing with high nonlinearity and features the advantage of providing the most relevant variables. Studies in predictions of energy consumption show such a strength when training with multi-dimensional complex data.[29] Predicting the patterns of energy consumption with the RF model actively incorporates the role of each parameter into the forecasting process, and therefore assists us to comprehend the main contributor of future energy changes. Given these advantages, RF is used in this work to investigate the non-linear relationship between energy consumption and influencing factors.

Hitherto the accuracy of total energy consumption forecasting remains highly uncertain,[30] improvement of which is increasingly important and urgent. In recent years, some predictions of global energy consumption have been made[11,31,32] by officials. However, these projections, though generated by large simulations, lacked detailed methods and a clear view of the main related contributors. Moreover, quantitative projection of future prospects of our civilization on the Kardashev Scale is nearly an open field. The pressing questions about the future of energy and mankind's progression on the Kardashev Scale require immediate answers with careful evaluation. Here, to provide a broad view of future energy developments before mankind enters a series of energy transformations and the projection of our civilization type, we use machine learning algorithms including RF and autoregressive integrated moving average[33] (ARIMA) to forecast the total energy consumption and the Kardashev index, K, of the human civilization from now to 2060. The projection presented in this study was made with a new forecasting strategy supported by completed datasets of a wide range of parameters. Through the generated results, a hypothesis is then tested for the potential advent of nuclear fusion, and we reach a conclusion of our civilization and energy developments for the near future.

**RESULTS**

This study adopts RF as the forecasting method, and the ARIMA model is incorporated to aid data deficiency (Figure 1). Eleven economic and climate parameters are considered in evaluating and forecasting the total energy consumption. Details about the model setup and data selection are in the Supplemental Information. Data for economic indicators and demographic parameters are provided by World Bank, including gross domestic product (GDP), population, urban population, and urbanization; data for climate variables are from Coupled Model Inter-comparison Project Phase 6 (CMIP6), including temperature, precipitation, carbon dioxide mole fraction, methane mole fraction, aerosol optical depth at 550nm, and atmosphere mass content of water vapor. We use the Shapley additive explanations[34] (SHAP) to estimate and illustrate the impact of each variable on the total energy consumption (Figure 2A). The SHAP value breaks down a projection to demonstrate the impact of each parameter through utilizing the traditional Shapley values, which is the weighted average of marginal contributions and their related extensions to connect optimal credit allocation with local explanations.[34] The RF model shows good performance. With 20% of the data extracted for validation (Figure 2B), it demonstrates high fidelity in projecting the observed energy consumption with $R^2$ and RMSE of 0.991 and 1.05, respectively.

Taking the advantage that RF provides a ranking of the variables based on their relative significance, the model concludes that GDP, total population, and urban population are the three most important influencing factors (Figure 2A). This indicates the strong connection between energy consumption and economy, though the causality between them is still being debated[35]. GDP shows a predominant role with a mean SHAP value of 4.63, exceeding that of the second most important parameter, population, by a factor of 6.4 and far greater than that of any other. This is expected as the growth of GDP usually demands energy consumption in almost all areas, and they show approximately a linear relationship (Figure S1). The world's total GDP has increased in the past few decades with an average of ~1.345 trillion dollars (2015 constant USD) each year (Figure S2B) from 18.05 trillion dollars in 1970 to 86.65 trillion dollars in 2021.[36] Also being economic indicators, population and urban population stand out as the second and the third largest contributors, likely due to their correlation with residential energy consumption. The global population has grown from 3.68 billion in 1970 to 7.76 billion (Figure S2C) in the past five decades, and it demonstrates a linear feature with an average annual increase of 81.4 million. Within the total population, the urban population shows a larger contribution than the rest, likely due to the fact that supporting urban activities, involving both industrial and residential usage of power, requires more energy. Results of the RF model indicate that energy consumption is closely related to economic and social factors, which will likely be the same case if current policy and technology trends continue.

Global energy consumption is predicted till 2060 using the trained RF model. A total of 42 countries that play major roles in the global economy and development, among the original 66 used for training, are selected in this forecast, and the results are further processed with the contribution ratio of these 42 countries to reach the conclusion. The central concern behind this approach is that the model is trained with data from individual countries; therefore, the prediction subject must correspond to its scale. Predictions of GDP, demographic variables, and climate variables are provided by the Organization for Economic Co-operation and Development, United Nations, and CMIP6, respectively. For the forecast of climate factors, several presumed shared socioeconomic pathway scenarios (SSP) are considered in CMIP6, which describe plausible alternative trends in the evolution of society and ecosystems over a century timescale. Individual scenarios are designated by the name of the basic pathway followed by two numbers indicating the increased radiative forcing achieved by the year 2100 in units of one-tenth watt.[37,38] For instance, the most optimized scenario, SSP126, is under the pathway of SSP1 (sustainability) with an additional radiative forcing of 2.6 W/m² by the end of the 21st century.

In order to achieve the total energy consumption prediction using the results of the selected 42 countries, we applied the ARIMA model to predict the proportion of the sum of these countries to the world's total (Figure 3). Given that the energy consumption growth of most selected countries follows their historical patterns and displays linear trends, time series modeling is an ideal tool to predict the said proportion. Results of the ARIMA show that such a ratio is projected to experience a gradual decline over the period of 2021 to 2060 from 0.773 to 0.731. Affected by the limitation of dataset size, the 95% confidence interval increases with time and reaches a relative uncertainty of 50% around 2045. Nonetheless, the said ratio experienced a steady decline over the past 25 years, and the prediction follows the same pattern.

Combining the energy assumption prediction of the selected countries and the forecasted trend of their portion, we achieved a prediction of the total energy consumption and mankind's position on the Kardashev Scale (Figure 4A, Table 1). The prediction shows that the world's total energy consumption is expected to reach 928-940 EJ ($9.28 \times 10^{20}$ J - $9.40 \times 10^{20}$ J) in 2060, with a growth of over 50% within the coming 40 years. Human civilization is prospected to achieve Type 0.7474 in 2060 with a humble average annual growth rate of approximately 0.077% from 2021 to 2060, and total growth of K=0.0198 from 2020. The energy consumption displays a linear feature in the prediction, which corresponds to a decreasing slope of the index K (Figure 4B). If there are no changes in the energy structure, the index K will likely increase even slower in the future.

Results of the models display non-linear relationships between energy consumption and other parameters. Presuming that the world does not experience major setbacks or technological breakthroughs, global energy consumption will likely follow the predicted path. However, new energy generation innovations may revolutionize the world energy sector. A scenario is tested assuming the advent of nuclear fusion around 2060, along with corresponding fast technological advancement and economic growth. Understanding the future in which nuclear fusion has the potential to revolutionize the world's energy and economy[39], we presume that its impact on humanity's development on the Kardashev Scale is identical to that of the most recent industrial revolution. The trend of K demonstrates a greater positive slope after the beginning of the Second Industrial Revolution (Figure S3), which indicates the exponential trend of energy consumption. More computational details are given in the Supplementary Information. Under this scenario, the total energy consumption is expected to experience rapid growth, leading the index K to increase and reach Type ~0.7719 at the end of the 21st century (Figure 4C), deviating from its current trend. In contrast, if the current energy structure does not change with a large composition of fossil fuels and no significant development of renewable technologies, human civilization is more likely to follow the trend colored in red, only be able to merely achieve Type ~0.7534 by the end of the 21st century.

**DISCUSSION**

The ability to harness, store, and consume energy reflects the development of a civilization, and the prediction of it provides instruction on constructing the future structure of world energy consumption, which potentially influences the future of our civilization.[40] The results of this study demonstrate a quantitative prediction of energy consumption. It fills the vacancy of mid-term projection of civilization type on the time scale of a few decades. Moreover, our results marked the first attempt to predict the future advancement of human civilization on the Kardashev Scale with multi-dimensional data taken into consideration. However, some associated elements that are not suitable to be quantified, such as national energy policy, may introduce additional uncertainties and challenges to the energy consumption prediction.[40,41] More accurate prediction is anticipated with updated data and advanced models. Nonetheless, our application of machine learning algorithms and in-depth data mining with comprehensive datasets maximized the precision and brought to light a harsh reality.

Our model concludes that total energy consumption is intimately linked with economic indicators. A noticeable phenomenon from the conclusion of the RF model is that environmental parameters show low impacts on energy consumption, which is potentially due to the fact they are not limiting factors at the current stage. However, such a situation may likely be subverted in the long term if the current energy structure continues with the portion of fossil fuels as high as 80%. The environmental parameters may become increasingly important in the future. Despite the said concern, the results indicate that our civilization will not be able to achieve significant progress on the Kardashev Scale in the next 40 years with the growth rate of K further declining. If the energy strategies and technologies remain in the present course, it may take human civilization millennia to become a Type 1 civilization. Prospective technologies such as carbon capture and sequestration, carbon neutrality, and nuclear fusion can potentially address this issue by either targeting sustainability or achieving significant energy production advancements.[42-45] A combination of them can bring civilization a bright future.

Presently, the development of mankind in the energy industry still unavoidably relies on fossil fuels, while many countries seek to fully substitute fossil fuels with renewable sources. A study proposed that the share of renewable energy can grow to 63% of the total primary energy supply in 2050, which can provide 94% of the greenhouse gas emissions reduction combined with higher energy efficiency.[46] As one of the most complex scientific and technical tasks, nuclear fusion has given humanity hope to generate a nearly inexhaustible, pollution-free source of energy.[47] Although the estimated time of the advent of nuclear fusion has kept changing for decades and the development of this technology is filled with uncertainties, its influence will be significant, and its position in the world's energy production is likely to be constructed soon after the technology becomes mature. An aggressive plan proposed by initiatives attempts to actively use fusion energy a few years after ITER's first plasma, which will potentially confirm fusion's role as a part of energy production in the second half of this century. As if right now, most researchers are foreseeing the actual deployment of nuclear fusion around 2060, confirming its unignorable role in future energy development.[15,48]

**CONCLUSION**

Energy is central to civilizations. The hidden patterns and behaviors of energy consumption and its related parameters provide clues for future energy growth. A well-analyzed prediction can be essential for humanity to realize its position on the Kardashev Scale and the world energy sector to make any necessary adjustments in favor of our development. Considering several contributing factors, we use the RF model to train the multi-dimensional data and make predictions based on it. Through a unique approach to aiding the results with ARIMA, we conclude that in 2060, human civilization will be able to achieve Type 0.7474 on the scale with an annual energy consumption of 928 EJ to 940 EJ under different shared socioeconomic pathways. Our methods significantly improve upon the previous projection of K, which oversimplifies the task. Uncertainties of the energy future and many existing concerns present a series of challenges to humanity, but technological innovations and sustainable approaches would lead to a promising future for our civilization.

## ACKNOWLEDGMENTS

This work is supported by the California Institute of Technology. Developers of the LightGBM package are gratefully acknowledged. We also thank Yuk L. Yung for the valuable advice, Yujun Xie for providing insights on nuclear fusion, and Kyle Sheng for his suggestions on the future of energy discussion.

## AUTHOR CONTRIBUTIONS

A.Z. conceived this study. J.Y. and A.Z. designed and performed the data processing and analytical methods. S. F. provided insights on interpreting and presenting the results. S.F. and J.Y. did supervision and project administration. All authors contributed to the writing and provided revisions for the manuscript.

## DECLARATION OF INTERESTS

The authors declare no competing interests.

## SUPPLEMENTAL INFORMATION

The analytical details and additional figure information are given in the Supplemental Information.

## LEAD CONTACT WEBSITE

https://www.gps.caltech.edu/people/jiani-yang
https://web.lmd.jussieu.fr/~sfan/

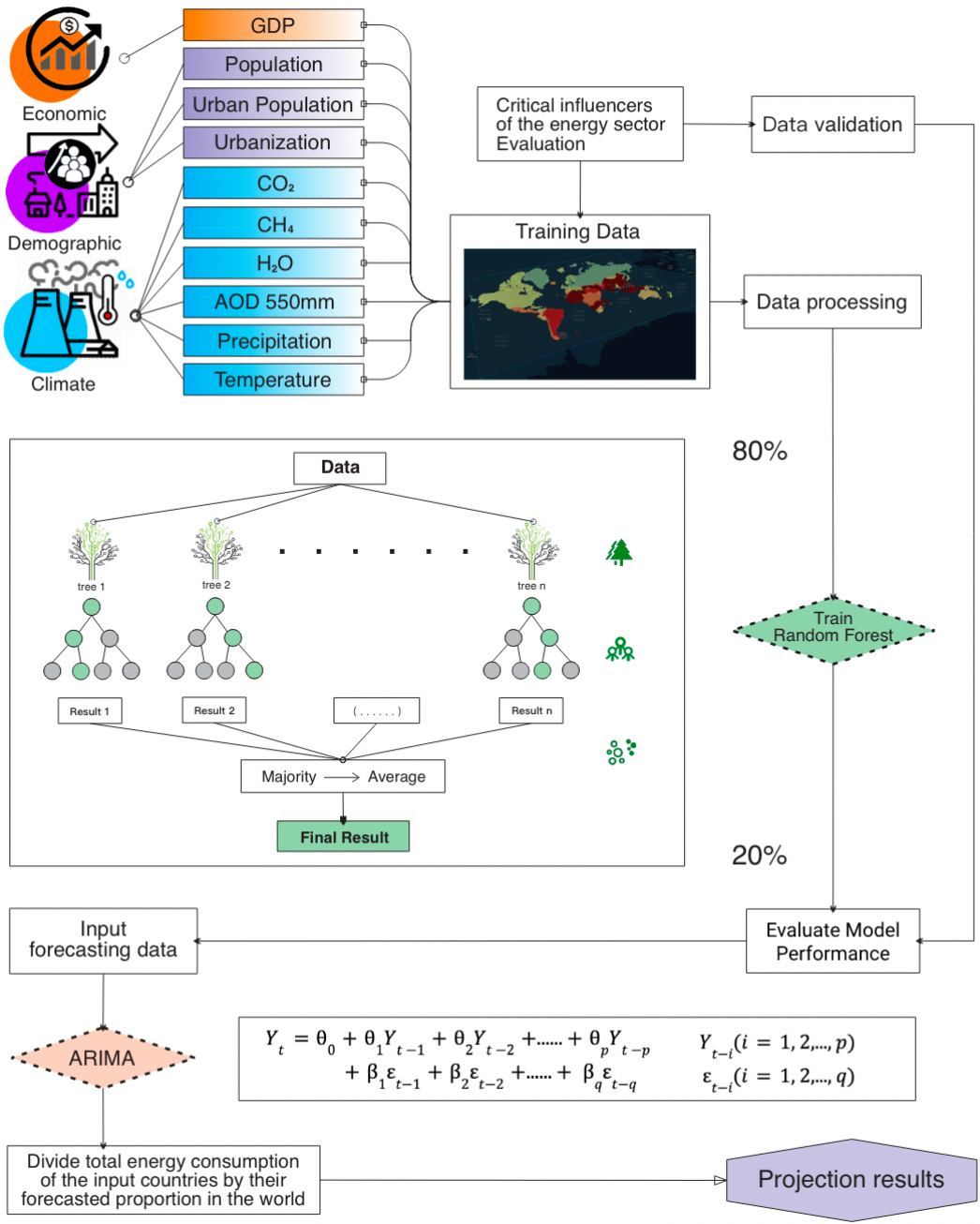

**Figure 1. Experiment procedures with variable charts and demonstration of applied machine learning models** The input parameters used to forecast energy consumption are listed and categorized. Each node represents a main step in the experiment procedures, and the arrows connect the steps. The RF model's general form is visualized and the equation for the ARIMA model is presented.

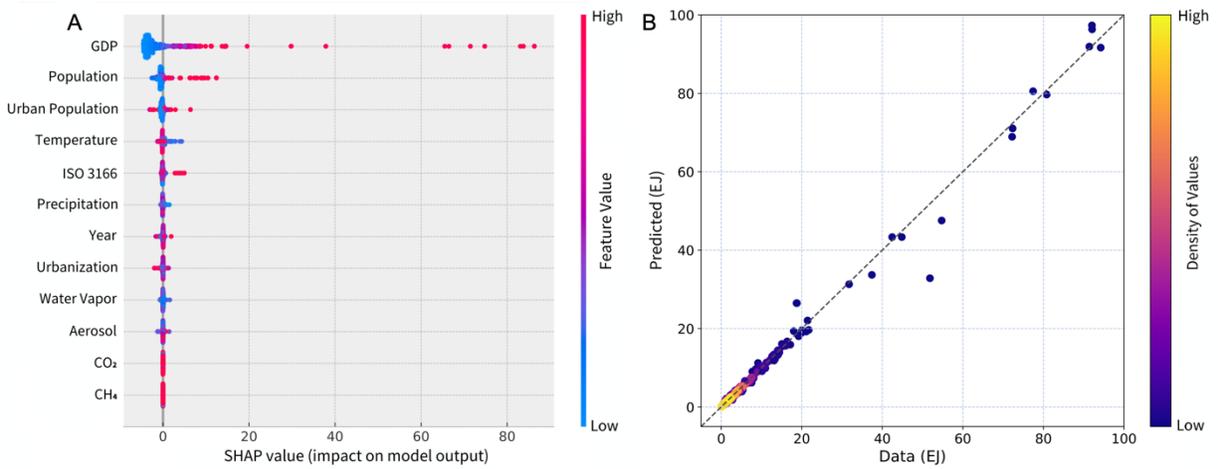

**Figure 2. Impact of major input parameters on the model output and performance of the RF model** (A) The selected factors' impacts on the total energy consumption are measured by their SHAP values. The color of each individual dot represents the value of that feature. Water vapor, aerosol, $CO_2$, and $CH_4$ are defined as atmosphere mass content of water vapor, aerosol optical depth at 550nm, carbon dioxide mole fraction, and methane mole fraction, respectively. (B) The model projection is compared with the validation dataset, which is extracted as 20% of the input data. A dashed line with a unit slope is shown for reference.

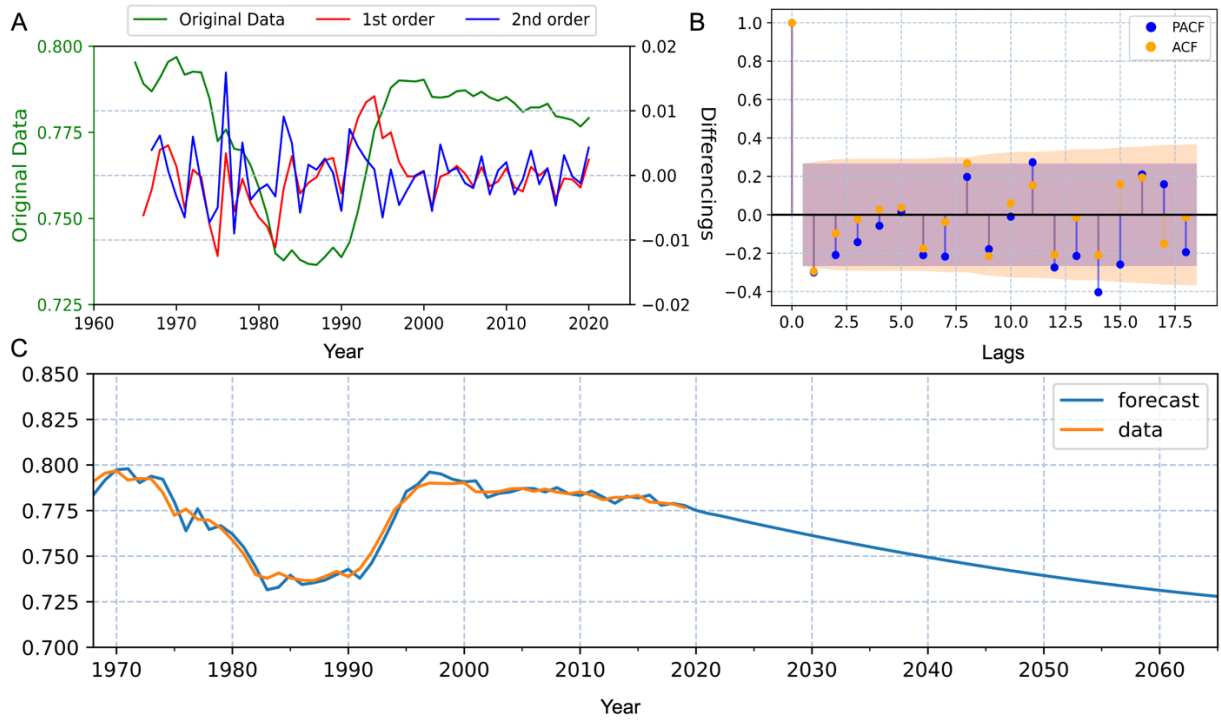

**Figure 3. Results of the ARIMA prediction** (A) Three orders of differencing. (B) Autocorrelation and partial autocorrelation of the model. (C) The ratio of the total energy consumption (yellow line) of the selected 42 countries to the global energy consumption together with the prediction results (blue line).

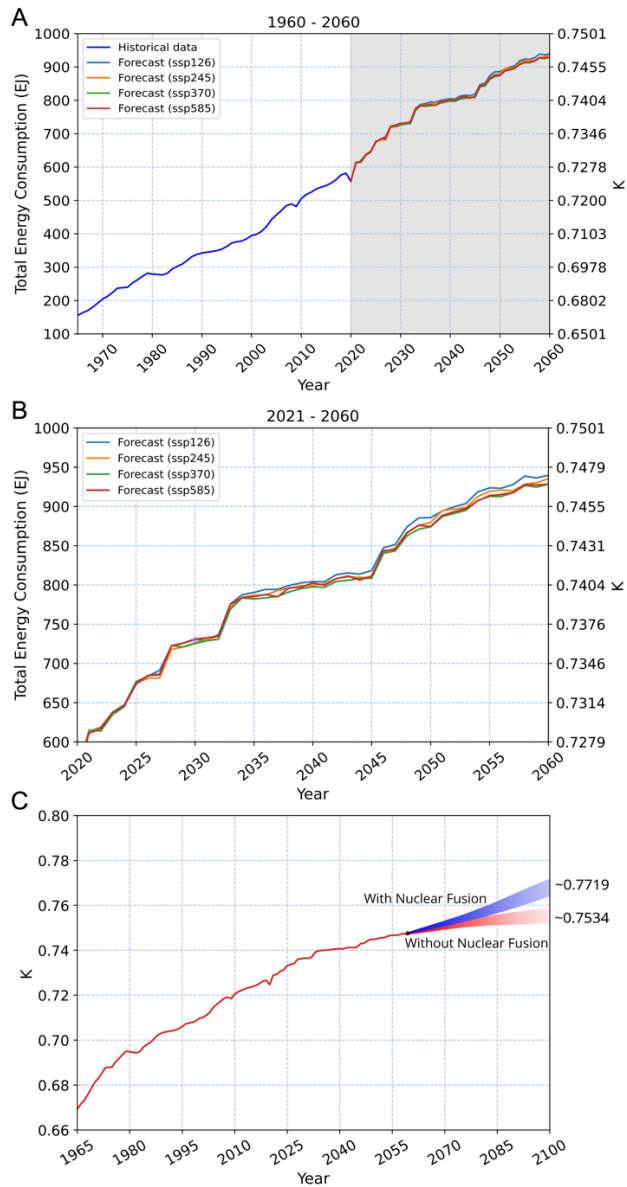

**Figure 4. Forecast of energy consumption and index K on the Kardashev Scale** Energy consumption is predicted through four shared socioeconomic pathway scenarios (SSP): SSP126 (light blue), SSP245 (yellow), SSP370 (green), and SSP585 (red). They each represent different socioeconomic scenarios with additional radioactive forcing of 2.6 W/m$^2$, 4.5 W/m$^2$, 7.0 W/m$^2$, and 8.5W/m$^2$, respectively. (B) is a zoom-in version of the shaded area of (A), focusing specifically on the predicted values. (C) The trend of the index K under two hypothetical scenarios, with (blue) and without (red) the advent of nuclear fusion in 2060, on mankind's development.

|      | Projected Values                                |                |                |                |         |                      |
| :---:| :---------------------------------------------: | :------------: | :------------: | :------------: | :-----: | :------------------: |
| Year | Energy Consumption (EJ)                         |                |                |                | K       |                      |
|      | SSP126                                          | SSP245         | SSP370         | SSP585         | Value   | Annual Growth Rate   |
| **2025** | 676.73                                      | 674.61         | 675.57         | 673.64         | 0.73316 | $2.64 \times 10^{-3}$ |
| **2030** | 730.34                                      | 726.24         | 725.38         | 731.11         | 0.73647 | $3.77 \times 10^{-4}$ |
| **2035** | 790.39                                      | 786.90         | 782.12         | 785.04         | 0.73990 | $2.15 \times 10^{-4}$ |
| **2040** | 804.26                                      | 799.43         | 797.57         | 802.43         | 0.74066 | $1.26 \times 10^{-4}$ |
| **2045** | 818.42                                      | 809.59         | 808.47         | 811.46         | 0.74142 | $3.32 \times 10^{-4}$ |
| **2050** | 885.83                                      | 879.66         | 874.05         | 874.33         | 0.74485 | $2.97 \times 10^{-5}$ |
| **2055** | 923.61                                      | 919.24         | 912.68         | 913.87         | 0.74667 | $3.17 \times 10^{-4}$ |
| **2060** | 939.72                                      | 935.51         | 928.46         | 928.16         | 0.74742 | $2.15 \times 10^{-4}$ |

**Table 1. Final forecasting results** The predicted value for energy consumption under four different presumed shared socioeconomic pathway scenarios, civilization development index K, as well as its growth rate from 2021 to 2060. See Table S1 for the full list of projected values.